\documentclass[12pt]{article}

\usepackage{wrapfig}
\usepackage{amssymb}
\usepackage{graphics}
\usepackage{wrapfig}
\usepackage{epsfig}
\usepackage{amsmath}
\usepackage{epsfig}

\begin{document}

\setcounter{page}{1}

\vspace{.5cm}

\begin{center}
{\Large \bf Effects of the Angular Momentum of Accreting Matter on the Flow Structure in the Subsonic Settling and Bondi-Hoyle Accretion Regimes}
\end{center}

\vspace{1cm}

\begin{center}
L.I. Arzamasskiy, V.S. Beskin\\
\vspace{1cm}
{\it Lebedev Physical Institute, Russian Academy of Sciences, Leninskii pr. 53, Moscow,~119991~Russia}\\
{\it Moscow Institute of Physics and Technology, Dolgoprudnyi,\\
Moscow oblast, 141700 Russia}\\
\end{center}

\noindent

\vspace{1cm}

Astronomy Letters, Volume 39, Issue 12, pp. 844-850 (2013)

Translated by G. Rudnitskii

\vspace{3cm}

{\small
\righthyphenmin=2

Previously unexplored accretion regimes associated with the rotation of accreting matter, namely the perturbations of a quasi-spherical subsonic settling flow and Bondi-Hoyle accretion in the presence of axial rotation, are considered within the framework of ideal hydrodynamics. For subsonic settling accretion, the perturbations are shown to grow rapidly as the gravitating center is approached, so that the flow in the inner regions can no longer be considered as quasi-spherical. For Bondi-Hoyle accretion, a vacuum cylindrical cavity is shown to be formed at large distances from the gravitating center near the flow axis, with the flow velocity outside this cavity being virtually independent of the distance to the rotation axis.
}
\noindent

\vspace{1cm}
Keywords: accretion, ideal hydrodynamics.

\newpage

\section{Introduction}

The accretion of matter onto a compact object (a neutron star onto which gas flows from the companion star in X-ray sources; a black hole that is the ``central engine'' in active galactic nuclei and quasars) is a classical problem of modern astrophysics (see Shapiro and Teukolsky 1983; Lipunov 1992; Bisnovatyi-Kogan 2011; and references therein). Beginning in the 1980s, the analytical approach whose foundation was laid back in the mid-twentieth century (Bondi and Hoyle 1944; Bondi 1952) began to be supplanted for natural reasons by numerical simulations (Hunt 1979; Petrich et al. 1989; Ruffert and Arnett 1994; Toropin et al. 1999; Toropina et al. 2012). Analytical solutions were found only in exceptional cases (Bisnovatyi-Kogan et al. 1979; Petrich et al. 1988; Anderson 1989; Beskin and Pidoprygora 1995; Beskin and Malyshkin 1996; Pariev 1996).

It should be emphasized that the focus of research has been shifted to magnetohydrodynamics, within which framework it has become possible to properly take into account the turbulent processes associated with magnetic reconnection, magnetorotational instability, etc. (Balbus and Hawley 1991; Brandenburg and Sokoloff 2002; Krolik and Hawley 2002). However, in our opinion, some of the important accretion regimes, which are simple enough for their main properties to be described analytically in terms of ideal hydrodynamics, still remain inadequately explored. These include the effects associated with the presence of angular momentum in the subsonic settling regime and for Bondi-Hoyle accretion. Such additional rotation naturally arises in binary systems when, for example, a neutron star interacts with the stellar wind from its companion, and when the gravitating center moves in a turbulent medium with significant vorticity. This paper is devoted to investigating such flows.

In the first part, we formulate the basic equations of ideal steady-state axisymmetric hydrodynamics, which are known to be reduced to one second-order equation for the stream function. Then, in the second part, the subsonic settling accretion is considered. We show that in the presence of angular momentum, the nonradial velocity perturbations grow fairly rapidly as the gravitating center is approached, so that the flow in the inner regions can no longer be considered quasi-spherical. Finally, the third part is devoted to the Bondi-Hoyle accretion. We show that in the presence of axial rotation, a vacuum cylindrical cavity is formed at large distances from the gravitating center near the flow axis. The flow velocity outside this cavity is virtually independent of the distance to the rotation axis.

\section{Basic equations}

As is well known (Guderley 1962; von Mises
1958), to describe axisymmetric steady hydrodynamic flows, it is convenient to use the stream function $\Phi(r,\theta)$ related to the poloidal velocity of the matter 
${{\bf v}}_{\rm p}$ by the following relation as an unknown quantity:
\begin{equation}
{{\bf v}}_{\rm p} = \frac{\nabla \Phi \times 
{{\bf e}}_{\varphi}}{2 \pi n_{\rm p} r_{\perp}},
\label{def}
\end{equation}
where $n_{\rm p}$ is the particle number density and $r_{\perp} = r \sin \theta$ is the cylindrical radius. At present, this approach is commonly called the method of the Grad-Shafranov equation (see, e.g., Beskin 2010). In this approach, the Euler equation is reduced to one second-order partial differential equation for the stream function $\Phi(r,\theta)$. In compact form (and for nonrelativistic flows, which will be considered below), it can be written as
\begin{equation}
-r_{\perp}^2 \nabla_k \left(\frac{1}{r_{\perp}^2 n_{\rm p}} \nabla^k \Phi \right) 
- 4 \pi^2 n_{\rm p} L \frac{{\rm d} L}{{\rm d} \Phi} + 4 \pi^2 r_{\perp}^2 n_{\rm p} 
\frac{{\rm d} E}{{\rm d} \Phi} - 4 \pi^2 r_{\perp}^2 n_{\rm p} \frac{T}{m_{\rm p}} 
\frac{{\rm d} s}{{\rm d} \Phi} = 0.
\label{main}
\end{equation}
Here, $m_{\rm p}$ is the mass of the particles and $T$ is the temperature. Equation (\ref{main}) 
represents the balance of forces in a direction perpendicular to the streamlines of the matter.

Next, the energy
\begin{equation}
E(\Phi) = \frac{v^2}{2} + w + \varphi_{\rm g},
\label{E}
\end{equation}
where $w$ is the enthalpy and $\varphi_{\rm g} = -GM/r$ is the
gravitational potential, the angular momentum
\begin{equation}
L(\Phi) = r_{\perp} v_{\varphi}
\label{L}
\end{equation}
and the entropy $s(\Phi)$ are integrals of motion, i.e., they are constant on streamlines and, consequently, may be considered as functions of  $\Phi$. Their specific form should be determined from the boundary conditions.

The Grad-Shafranov equation (\ref{main}) should be supplemented with the Bernoulli equation (\ref{E}), which can now be rewritten as
\begin{equation}
\frac{(\nabla \Phi)^2}{8 \pi^2 n_{\rm p}^2 r_{\perp}^2} + \frac{L^2}{2r_{\perp}^2} + w - \frac{GM}{r} = E.
\label{Bernulli}
\end{equation}

Below, for simplicity, we will use the polytropic
equation of state
\begin{equation}
P=k(s)n_{\rm p}^{\Gamma},
\end{equation}
where $\Gamma$ is the polytropic index, and $k(s)$ depends only on the entropy $s$. In this case, the speed of sound is a function of the number density $n_{\rm p}$ and can be expressed as
\begin{equation}
c_{\rm s}^2 = \frac{1}{m_{\rm p}} \Gamma k(s) n_{\rm p}^{\Gamma-1}.
\end{equation}
Accordingly, at $\Gamma \neq 1$ the enthalpy can be represented as
\begin{equation}
w = \frac{c_{\rm s}^2}{\Gamma - 1}.
\label{wdet}
\end{equation}

\section{Subsonic settling accretion}

Consider the problem of subsonic settling accretion of matter. This regime corresponds to a subsonic flow up to the gravitating center. In this case, the contribution from the first term associated with the kinetic energy of the matter in the Bernoulli equation (\ref{E})
is assumed to be small, so that in the inner regions we can set
\begin{equation}
\frac{c_{\rm s}^2}{\Gamma - 1} \approx \frac{GM}{r}.
\label{slow}
\end{equation}

If the angular momentum of the accreting matter is small enough, so that  $v_{\varphi} \ll v_{\rm p}$
in the entire flow region, then it is natural to assume that the flow structure will differ only slightly from the spherically symmetric case. Therefore, we can seek a solution of our problem in the form
\begin{equation}
\Phi(r,\theta) = \Phi_0 [1-\cos \theta + \varepsilon_L^2 f(r,\theta)],
\end{equation}
where the last term is the correction to the spherically symmetric flow. As regards the small parameter $\varepsilon_L^2$, it will be determined somewhat later.

Let us now determine the conditions at the outer boundary of the flow. Since the flow is assumed to be subsonic, we will need five boundary conditions. Three of them, namely two thermodynamic functions and radial velocity $v_{r}$ can be chosen from the zeroth spherically symmetric approximation. Therefore, at the outer boundary $r = R$ 
we set
\begin{eqnarray}
T(R, \theta) & = & T_{R}, \\
n_{\rm p}(R, \theta) & = & n_{R}, \\
v_{r}(R, \theta) & = & v_{R}.
\end{eqnarray}
Below, we will assume that in the presence of slow rotation, the temperature $T_{R}$, 
number density $n_{\rm R}$ and meridional velocity $v_{R}$ at $r = R$ do not change and the gas rotates as a whole with an angular velocity $\Omega$. In this case, we can write
\begin{eqnarray}
v_{\varphi}(R, \theta) & = &  \Omega R \sin\theta, 
\label{bc4}\\
v_{\theta}(R, \theta) & = & 0.
\label{bc5}
\end{eqnarray}
Then,
\begin{eqnarray}
L(\Phi) & = & R v_{\varphi} \sin \theta = L_0 \sin^2 \theta, \\
E(\Phi) & = & E_0 + \frac{1}{2} m_{\rm p} v_{\varphi}^2 = E_0+\frac{L_0^2}{2R^2} \sin^2 \theta,
\end{eqnarray}
where $L_0 = \Omega R^2$ and $E_0$ is the value of the Bernoulli
integral in the absence of rotation. 
Accordingly, the total accretion rate $\Phi_{\rm tot} = 2 \Phi_0$ will be
\begin{equation}
\Phi_{\rm tot} = 4 \pi m_{\rm p} n_{R} v_{R}R^2.
\end{equation}

As a result, after linearization in small parameter $\varepsilon_L$ and in the limit of low velocities $v_{\rm p} \ll c_{\rm s}$, equation \eqref{main} can be rewritten as
\begin{align}
-\varepsilon_L^2\left[\frac{\partial^2 f}{\partial r^2} 
+ \frac{\sin \theta}{r^2} \frac{\partial}{\partial \theta} 
\left( \frac{1}{\sin \theta} \frac{\partial f}{\partial \theta} \right) 
+ \frac{GM}{r^2 c_{0}^2} \frac{\partial f}{\partial r}\right] = 
%\nonumber \\
\frac{L_0^2}{r^2 v_{0}^2} \left( \frac{2}{r^2} 
- \frac{1}{R^2} \right) \sin^2 \theta \cos \theta\label{eqn1},
\end{align}
where $v_{0}(r)$ and $c_{0}(r)$ are the poloidal velocity and the speed of sound for an unperturbed spherically symmetric flow. As we see, the small parameter of our problem is
\begin{equation}
\varepsilon_L = \frac{\Omega R}{v_{R}}.
\label{sp1}
\end{equation}

Expanding now the function $f(r,\theta)$ into a series of eigenfunctions $Q_m(\theta)$ of the operator
%\begin{equation}
$\hat{\mathcal{L}}_\theta = \sin \theta \, \partial/\partial \theta 
\left[( 1/\sin \theta) \partial /\partial \theta \right]$ 
%\end{equation}
\begin{equation}
f(r, \theta) = \sum_{m=0}^{\infty}g_m(r)Q_m(\theta),
\end{equation} 
where (see Beskin 2010) $Q_0 = (1 - \cos\theta)$, $Q_1 = \sin^2\theta$, $Q_2 = \sin^2\theta\cos\theta$,
etc., and substituting this series into equation \eqref{eqn1}, we obtain a system of ordinary differential equations for the radial functions $g_m(r)$:
\begin{equation}
-r^2 \frac{{\rm d}^2 g_m}{{\rm d} r^2} - \frac{1}{(\Gamma - 1)}r 
\frac{{\rm d} g_m}{{\rm d} r} - q_{m} g_m = 0, \quad m \ne 2,
\end{equation}
\begin{align}
-r^2 \frac{{\rm d}^2 g_m}{{\rm d} r^2} - \frac{1}{(\Gamma - 1)}r  
\frac{{\rm d} g_m}{{\rm d} r} + 6 g_m %= \nonumber \\
=  \frac{v_{R}^2}{v_{0}^2} 
\left(2\frac{R^2}{r^2} - 1 \right), \quad m=2.
\label{eqn2}
\end{align}
Here, $q_m = -m(m+1)$ are the eigenvalues of the
operator $\hat{\mathcal{L}}_\theta$
and we also used equation (\ref{slow}). Using now definition (\ref{def}) and boundary conditions (\ref{bc4})--(\ref{bc5}), we find that all radial functions should satisfy the conditions
\begin{eqnarray}
g_m(R) & = & 0, 
\label{bc1} \\
\left.\frac{{\rm d}g_m}{{\rm d}r}\right|_{r = R} & = & 0.
\label{bc2}
\end{eqnarray}
As a result, only the radial function  $g_2(r)$,
turns out to be nonzero. For the latter, equation \eqref{eqn2} takes the form

\begin{equation}
r^2 g_2'' + \frac{1}{\Gamma-1} r g_2' - 6 g_2 
= -2 K \cdot \left(\frac{r}{R}\right)^{\alpha} + K \cdot \left(\frac{r}{R}\right)^{\alpha+2},
\label{en}
\end{equation}
where
\begin{equation}
K=  (\Gamma - 1)^{2/(\Gamma-1)} \left(\frac{GM}{c_{R}^{2}R}\right)^{2/(\Gamma-1)}
\end{equation}
and
\begin{equation}
\alpha = \frac{2(\Gamma-2)}{\Gamma-1}.
\end{equation}
As a result, the solution of equation (\ref{en}) with boundary conditions (\ref{bc1})--(\ref{bc2})
can be represented as
\begin{equation}
g_2 = 
C_1 \left(\frac{r}{R}\right)^{\lambda_1}
+ C_2 \left(\frac{r}{R}\right)^{\lambda_2}
+ K_1 \left(\frac{r}{R}\right)^{\alpha} 
+ K_2\left(\frac{r}{R}\right)^{\alpha+2},
\end{equation}
where 
\begin{equation}
\lambda_{1,2} = \frac{(\Gamma-2) \pm \sqrt{(\Gamma-2)^2 + 24(\Gamma-1)^2}}{2(\Gamma-1)}
\end{equation}
and
\begin{eqnarray}
C_1 & = & K \frac{\alpha+4-\lambda_1}{(\lambda_1-\lambda_2)(\alpha-\lambda_1)(\alpha+2-\lambda_2)}, \\
C_2 & = & - K \frac{\alpha+4-\lambda_2}{(\lambda_1-\lambda_2)(\alpha-\lambda_2)(\alpha+2-\lambda_1)}, \\
K_1 & = & - \frac{2K}{(\alpha-\lambda_1)(\alpha-\lambda_2)}, \\
K_2 & = & \frac{K}{(\alpha+2-\lambda_1)(\alpha+2-\lambda_2)}.
\end{eqnarray}

\begin{figure}[h!]
\includegraphics[scale=1]{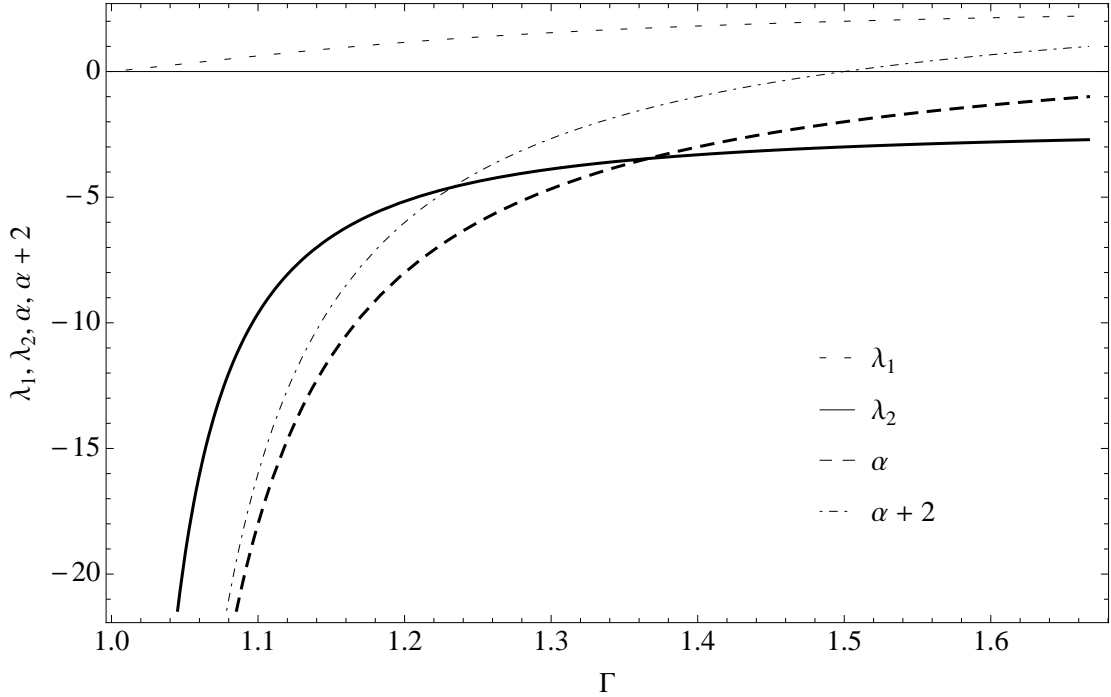}
\centering
  \caption{Exponents  $\lambda_1$, $\lambda_2$ and $\alpha$ versus $\Gamma$. The leading exponents determining the growth of perturbations as the gravitating center is approached are highlighted.}
  \centering
\label{Fig1}
\end{figure}

\begin{table}[h!]
\centering
\begin{tabular}{|c|c|c|c|c|c|c|c|c|c|}
\hline
$\Gamma$ & 1.01 & 1.1 & 1.2 & 4/3 & 1.366 & 1.4 & 1.5 & 1.6 & 5/3 \\
\hline
 $\lambda_1$ & 0.06 & 0.62 & 1.16& 1.65 & 1.73 & 1.81& 2.00 & 2.14 & 2.21 \\
\hline
$\lambda_2 $& $-99.1$ & $-9.62$ & $-5.16$ & $-3.65$ & $-3.46$ & $-3.31$ & $-3.00$ & $-2.81$ & $-2.71$  \\
\hline
$\alpha$ &$ -198$ & $-18.0$ & $-8.00$ & $-4.00$ & $-3.46$ & $-3.00$ & $-2.00$ & $-1.33$ & $-1.00$ \\
\hline
\end{tabular}
  \caption{Dependence of the exponents $\lambda_1$, $\lambda_2$ and $\alpha$ on $\Gamma$. }
\label{Tab2}
\end{table}

In Figure \ref{Fig1}, the exponents $\lambda_1$, $\lambda_2$ and $\alpha$ 
are plotted
against the polytropic index $\Gamma$; the numeric values corresponding to the plot are presented in the table. As we see, at $\Gamma = 1.366$ the leading exponents change from $\lambda_2$ to $\alpha$. However, at all values of $\Gamma < 5/3$ the leading exponents are negative and greater than 2.7 in magnitude. Therefore, at small radii r, the radial flow perturbations should grow rapidly.
 
For instance, in the case considered by Shakura et al. (2013), the outer region of the flow is determined by the capture of the stellar wind by a neutron star in a binary system. In this case, the parameter $K$ turns out to be of the order of unity, and the lower boundary of the subsonic settling flow (determined by the size of the magnetosphere) will be smaller than the outer radius of the flow $R$ by two or three orders of magnitude. Since the rotation velocity in this paper is estimated to be several percent of the Keplerian
velocity, $v_{\varphi}\approx 0.05 (GM/r)^{1/2}$, and since the radial velocity $v_{R}$ for the subsonic settling regime should be, by definition, smaller than the free fall velocity, the parameter $\varepsilon_{L} = v_{\varphi}/v_{R}$
(\ref{sp1}) turns out to be of the order of unity. On the other hand, the perturbations increase approximately by a factor of 1000 already at distances that are only a factor of 10 smaller than $R$. Therefore, even for a small parameter
$\varepsilon_L^2$  of $\sim 0.01$ in the inner regions, we can no longer consider the perturbations
$\varepsilon_L^2 g_2(r)\sin^2\theta\cos\theta$ to be small and the accretion to be quasi-spherical.

\section{Bondi-Hoyle accretion in the presense of axial rotation}

Let us now consider another classical example, namely the Bondi-Hoyle accretion; in this case, it is natural to pass to the frame of reference in which the gravitating center is at rest. The method of the Grad-Shafranov equation considered here allows us to analyze the flows with accreting matter rotating around the axis along which the gravitating center moves. Just as in Beskin and Pidoprygora (1995), we will consider the case of subsonic motion, $v_{\infty} \ll c_{\infty}$, where $v_{\infty}$ s the velocity of the gravitating center and $c_{\infty}$ is the sound speed for the medium at infinity. So the first small parameter of the problem will be their ratio
\begin{equation}
\varepsilon_1 = \frac{v_{\infty}}{c_{\infty}},
\label{eps1}
\end{equation}
As is well known, the presence of small parameter (\ref{eps1}), allows an analytical solution of the problem to be constructed (Beskin 2010). In particular, the capture radius $R_{\rm c}$ can be defined as
\begin{equation}
R_{\rm c} \approx r_{*} \varepsilon_{1}^{-1/2},
\label{Rc}
\end{equation}
where
\begin{equation}
r_{\rm *} = \frac{(5 - 3\Gamma)}{4}\frac{GM}{c_{\infty}^2}
\label{r*}
\end{equation}
is the sonic surface.

Let us now assume that the incoming flow has a small angular momentum $L$. Clearly, significant perturbations will be concentrated only near the rotation axis, because the streamlines with an angular momentum $L$ in this accretion regime cannot approach
the rotation axis closer than the distance.
\begin{equation}
r = \sqrt{\frac{L^2}{2 E}}.
\end{equation}
Consequently, we will be interested only in the regions
near the separatrix separating the captured streamlines and the streamlines going to infinity. Therefore,
for simplicity, we will assume that all three integrals of
motion, $E(\Phi)$, $L(\Phi)$ and $s(\Phi)$, are constants near the
separatrix. For instance, for the angular momentum $L$ this implies that $v_{\varphi} \propto r_{\perp}^{-1}$ in the onflow region (and
only near the separatrix). In this case, equation
\eqref{main} takes a particularly simple form:
\begin{equation}
-r_{\perp}^2 \nabla_k \left(\frac{1}{r_{\perp}^2 n_{\rm p}}\nabla^k \Phi \right) =0.
\label{point}  
\end{equation}
Another small parameter of the problem will be
\begin{equation}
\varepsilon_2 = \frac{v_{\varphi}(R_{\rm c})}{c_{\infty}},
\label{eps2}
\end{equation}
where the toroidal velocity $v_{\varphi}$ is taken on the separatrix in the onflow region, i.e., at \mbox{$r_{\perp} = R_{\rm c}$.} As a result, the integrals of motion
$E$ and $L$ near the separatrix will be
\begin{eqnarray}
E & = & \frac{c_{\infty}^2}{\Gamma - 1}, \\
L & = & \varepsilon_{2} R_{\rm c}c_{\infty}.
\end{eqnarray}

Clearly, in contrast to the previous problem, the separation of variables here is impossible. Therefore, we will restrict our analysis only to the flow structure near the rotation axis, where the influence of a small angular momentum turns out to be significant. First of all, consider the asymptotic region $r \rightarrow \infty$ along the flow, where the gravitational potential may be set equal to zero. In this case, the Bernoulli equation can be written as
\begin{equation}
\frac{L^2}{2 r_{\perp}^2} + \frac{\Gamma k(s)n_{\rm p}^{\Gamma - 1}}{\Gamma - 1} 
= \frac{\Gamma k(s)n_{\infty}^{\Gamma - 1}}{\Gamma - 1}.
\label{Bern2}
\end{equation}
On the other hand, equation (\ref{point}) will be rewritten as
\begin{equation}
r_{\perp} \frac{{\rm d}}{{\rm d}r_{\perp}}
\left[\frac{1}{r_{\perp}n_{\rm p}(r_{\perp})}\frac{{\rm d}\Phi}{{\rm d}r_{\perp}}\right] = 0.
\label{GSGS}
\end{equation}
Since we may set $\Phi = \pi r_{\perp}^2 n_{\infty} v_{\infty}$ when $r_{\perp} \rightarrow \infty$ we obtain
\begin{equation}
\frac{{\rm d}\Phi}{{\rm d}r_{\perp}} = 2 \pi r_{\perp}
\left[1 - \left(\frac{r_{\perp}}{r_{\rm min}}\right)^{-2}\right]^{1/(\Gamma - 1)} v_{\infty}n_{\infty},
\end{equation}
where 
\begin{equation}
r_{\rm min} = \varepsilon_{2}\left(\frac{\Gamma - 1}{2}\right)^{1/2}R_{c}
\label{rmin}
\end{equation}
is the minimum distance to which the flow can approach the rotation axis.

On the other hand, neglecting the derivatives with respect to $z$ and using equation (\ref{point}),
we conclude that
\begin{equation}
v_{\rm p} = v_z = {\rm const},
\end{equation}
i.e., the poloidal velocity of the flow does not depend on the distance to the axis. The value of 
$v_{\rm p}$ itself in the asymptotically distant region should coincide with the onflow velocity $v_{\infty}$. 
Finally, from the Bernoulli integral (\ref{Bern2}) we obtain
\begin{equation}
n_{\rm p} =
\left[1 - \left(\frac{r_{\perp}}{r_{\rm min}}\right)^{-2}\right]^{1/(\Gamma - 1)}n_{\infty},
\label{nnn}
\end{equation}
i.e., the number density grows with distance from the
axis and reaches a constant level.

\begin{figure}[h!]
\includegraphics[scale=0.25]{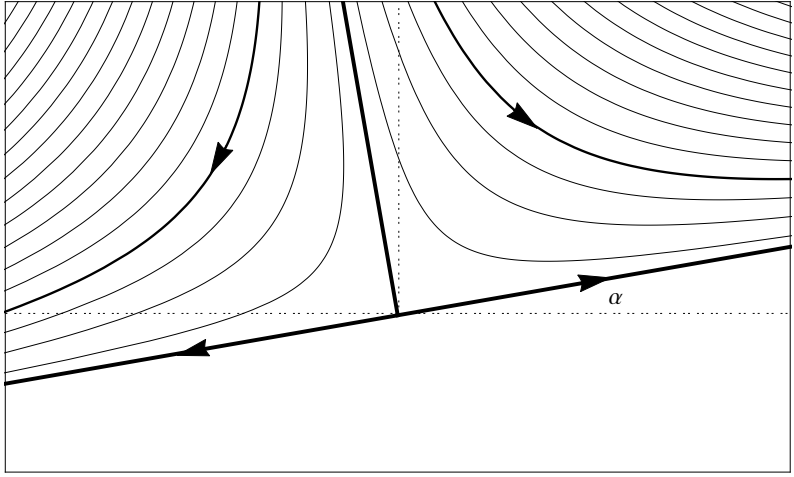}
\centering
  \caption{Flow structure near the singular point. There is a vacuum cavity at small distances from the axis, while the flow separates into two parts at large distances. The matter to the left of the separatrix moves in one direction and to the right in the other direction.}
  \centering
\label{sp}
\end{figure}

Let us now consider the flow structure near the singular point (i.e., the point separating the flows going to infinity and returning to the gravitating center; see Figure \ref{sp}). Near this point, the Grad-Shafranov equation will still be determined by equation (\ref{point}). As regards the Bernoulli equation, in this case, it takes the form
\begin{equation}
\frac{(\nabla \Phi)^2}{8 \pi^2 r_{\perp}^2 n_{\rm p}^2} + \frac{L^2}{2 r_{\perp}^2} 
+ \frac{c_{\rm s}^2}{\Gamma-1} - \frac{G M}{r} = E.\label{bernulli} 
\end{equation}

Clearly, both the number density $n_{\rm p}$ and the poloidal velocity $v_{\rm p}$ must become zero near the singular point. Therefore, we will seek a solution of equation \eqref{point} in the form
\begin{eqnarray}
n_{\rm p}(r_{\perp}, z) & = & A [r_{\perp}-r_{0} (z)]^n; \\
\Phi(r_{\perp}, z) & = & B [r_{\perp}-r_{0} (z)]^m [z-z_0 (r_{\perp})],
\end{eqnarray}
where $r_{0}$ and $z_{0}$ are the coordinates of the singular point, $A$ and $B$ are some dimensional constants, $m$ and $n$ are the exponents that must be greater than zero. In this case, it is natural to assume
$z_0$ to be much greater than $r_{0}$. Substituting these functions into
\eqref{point}, 
we obtain
\begin{align}
[r_{\perp}-r_{0}(z)]^{m-n-2} [z-z_0 (r_{\perp})]\cdot \left[ 
m(m-n-1) - n m r_{0} '(z) + m(m-1) (r_{0} ' (z))^2\frac{}{}  \right] + 
\nonumber \\
[r_{\perp}-r_{0}(z)]^{m-n-1} \cdot \left[-m/r_{\perp} \cdot 
(z-z_0(r_{\perp})) + n z_0'(r_{\perp}) -2 m z_0'(r_{\perp})  \right.
\nonumber\\
\left.+ nr_{0}'(z) - 2m r_{0}'(z) - m [z-z_0(r_{\perp})] r_{0}''(z)  \right]+
[r_{\perp}-r_{0}(z)]^{m-n} \cdot \left[  \frac{z_0'(r_{\perp})}{r_{\perp}} 
- z_0''(r_{\perp}) \right] = 0. 
\label{sum}
%(r_{\perp}-r_{0}(z))^{m-n} \cdot \left[  \frac{z_0'(r_{\perp})}{r_{\perp}} 
%- z_0''(r_{\perp}) \right] = 0. 
%\nonumber
\end{align}

Near the singular point, i.e., for $r_{\perp} \rightarrow r_{0} (z)$ and
$z \rightarrow z_0(r_{\perp})$, only the first two terms will contribute significantly to sum \eqref{sum}.  Therefore, we can write
\begin{equation}
m(m-n-1) + m(m-n-1) (r_{0} ' (z))^2 =0,
\label{35}
\end{equation}
\begin{equation}
-\frac{m (z-z_0(r_{\perp}))}{r_{\perp}} + n z_0'(r_{\perp}) -2 m z_0'(r_{\perp}) 
+ nr_{0}'(z)- 2m r_{0}'(z) - m (z-z_0(r_{\perp})) r_{0}''(z) =0,
\label{36}
\end{equation}
where the primes denote the derivatives with respect to the corresponding argument. As a result, equation \eqref{35} gives
\begin{equation}
m-n-1=0.
\label{mn}
\end{equation}
At the same time, the condition of the first order in \eqref{36} can be rewritten as
\begin{equation}
r_{0}'(z) + z_0'(r_{\perp}) = 0.
\end{equation}
This is equivalent the lines $r_{0}(z)$ and $z_0 (r_{\perp})$ being perpendicular to each other. Therefore, near the singular point, we may set
\begin{eqnarray}
r_{0}(z) & = & r_{0} + \alpha (z - z_{0}), \\
z_{0}(r_{\perp}) & = & z_{0} - \alpha (r_{\perp} - r_{0}).
\end{eqnarray}

Let us now analyze the Bernoulli equation. For
convenience, we introduce the notation
\begin{align}
\Delta r_{\perp}&=r_{\perp} - r_{0} (z), \\
\Delta z&=z - z_0 (r_{\perp}).
\end{align}
For $r_{\perp}$ and $z$, we can then derive the expressions
\begin{equation}
r_{\perp} = r_{0} + \frac{\Delta r_{\perp} + \alpha \Delta z}{1 + \alpha^2},
\end{equation}
\begin{equation}
z = z_0 + \frac{\Delta z - \alpha \Delta r_{\perp}}{1 + \alpha^2}.
\end{equation}
Considering now the terms of the zeroth and first order in $\Delta r_{\perp}$ and $\Delta z$ in equation \eqref{bernulli}, we obtain
\begin{equation}
\frac{L^2}{2 r_{0}^2} -\frac{G M}{ z_0} = \frac{c_{\infty}^2}{\Gamma - 1},
\label{1}
\end{equation}
\begin{equation}
-\frac{\alpha L^2}{r_{0}^3} + \frac{G M}{z_0^2} = 0,
\label{2}
\end{equation}
\begin{equation}
- \frac{G M}{\alpha z_{0}^2}
+ k(s) \frac{1}{m_{\rm p}} \frac{\Gamma}{\Gamma-1} A^{\Gamma-1} = 0.
\label{3}
\end{equation}

First of all, is is easy to show that the contribution of the gravitational term in equation (\ref{1}) has the order of
smallness $\varepsilon_{1}^{1/2}$ and we will neglect it below. As a
result,  (\ref{2}) gives
\begin{equation}
\alpha \approx \varepsilon_{1}^{1/2}\varepsilon_{2}.
\label{allph}
\end{equation}
Therefore, the inclination of the boundary $r_{0}(z)$ to the flow axis turns out to be very small. The coordinate of the singular point $r_0$ will differ only slightly from 
$r_{\min}$ (\ref{rmin}). Next, from equation (\ref{3}) we find
\begin{equation}
A \approx n_{\infty} r_{*}^{-1/{(\Gamma - 1)}}.
\label{Aph}
\end{equation}
Finally, the terms in equation \eqref{3} will correspond to the coefficients at the identical powers of $\Delta z$ only under the condition
\begin{equation}
n = \frac{1}{\Gamma-1}.
\label{4}
\end{equation}
We see that, given equation (\ref{mn}) the quantity $A$ and the exponent
$n$ are found in accordance with equation (\ref{nnn}). The flow structure near the singular point can be understood from Figure~\ref{sp}.

In conclusion, we will provide the expressions for the innermost, supersonic flow regions $r \ll r_{*} \approx  GM/c_{\infty}^2$ without their derivation. For sufficiently small angular momenta, when the accreting plasma cannot penetrate only into a narrow quasi-cylindrical region with a distance from the axis $r_{\rm min} \ll r$, the quantity 
$r_{\rm min}$ can be represented as
\begin{equation}
r_{\rm min} \approx \varepsilon_{2}\varepsilon_{1}^{-1/2}
%2^{(\Gamma-3)/2}(5 - 3\Gamma)^{1/2}(\Gamma - 1)^{1/2}
r_{*}\left(\frac{r}{r_{*}}\right)^{3(\Gamma - 1)/4}.
\label{rinmin}
\end{equation}
As we see, such a solution can be realized only under the condition
$\varepsilon_{2} < \varepsilon_{1}^{1/2}$. Otherwise, in the case of
ideal hydrodynamics we consider, the flow will be unable to approach the rotation axis to a distance smaller the radius of the sonic surface $r_{*}$. 

Further, for $r_{\rm min} \ll r_{\perp} < r$ , the solution should coincide with the unperturbed (i.e., essentially cylindrical) supersonic flow with an accretion rate
$v_{\rm p}^{(0)}(r) \approx (2GM/r)^{1/2}$. Owing to equation (\ref{GSGS}), the poloidal velocity $v_{\rm p}(r, r_{\perp})$, just as for the outgoing flow, will not depend
on the distance from the axis $r_{\perp}$. Accordingly, the
particle number density $n_{\rm p}$  
will be given by a relation
similar to  (\ref{nnn}) 
\begin{equation}
n_{\rm p} (r, r_{\perp}) =
\left[1 - \left(\frac{r_{\perp}}{r_{\rm min}}\right)^{-2}\right]^{1/(\Gamma - 1)}
n_{\infty}(r),
\label{nnnew}
\end{equation}
where now, however, the particle number density $n_{\infty}(r) \approx n_{*}(r/r_{*})^{-3/2}$
depends on the distance to the gravitating center.

As an illustration, we will provide estimates for the case where the compact object (a neutron star, a black hole) moves through a turbulent cell of an interstellar cloud with a temperature of $100$--$1000^{\circ}$ K. The sound speed in them will then be approximately $10^4$--$10^5$ cm/s. At the same time, the tangential velocity of the matter at the boundary for a typical cloud is $10^6$ cm/s (Horedt 1982). If the cloud velocity toward the gravitating center is taken to be $10$ km/s, then the parameters of the problem will be
\begin{equation}
\varepsilon_1 \sim 1;~\varepsilon_2  \sim 10.
\end{equation}
As we see, in this case, these parameters can be fairly large. And this implies that the flow structure can actually change significantly when the accreting matter has an angular momentum. For instance, at  $\varepsilon_1=\varepsilon_2=1$ the capture radius turns out to be equal to the distance to the sonic surface and the radius of the vacuum cavity coincides with this quantity in order of magnitude. Clearly, a consistent comparison of the predictions of the theory and observations requires a separate study, which is beyond the scope of this paper.

\section{Conclusions}

Thus, as has been shown above, the presence of even a small angular momentum in the accreting matter can qualitatively change the flow structure in both subsonic settling and Bondi-Hoyle accretion regimes. Therefore, the effects associated with the rotation of accreting matter can affect significantly the overall picture of the phenomena being discussed when analyzing accretion onto compact astrophysical objects.

Of course, the simple solutions of the equations of ideal hydrodynamics considered above cannot describe the entire spectrum of phenomena associated with the axial rotation of accreting matter and attributable, for example, to turbulence or a magnetic field. Using the polytropic equation of state at gas densities approaching zero is also a weak point of the approach considered. Nevertheless, in our view, the simple examples discussed above can be used as a first step that allows us to judge how the rotation of accreting matter changes the flow structure in the well-known cases of subsonic settling or Bondi-Hoyle accretion.

We wish to thank G.M. Beskin, K.A. Postnov, and A.A. Philippov for a helpful discussion. This work was supported by the Federal Goal-Oriented Program of the Ministry of Education and Science, contracts nos. 14.A18.21.0790 and 14.V37.21.0915.

\begin{center}
REFERENCES
\end{center}
\begin{enumerate}
\item M. Anderson, Mon. Not. R. Astron. Soc. {\bf 239}, 19 (1989).
\item S. A. Balbus and J. F. Hawley, Astrophys. J. {\bf 376}, 214 (1991).
\item V. S. Beskin, MHD Flows in Compact Astrophysical Objects (Springer, Berlin, 2010; Fizmatlit, Moscow, 2005).
\item V. S. Beskin and L. M. Malyshkin, Astron. Lett.  {\bf 22}, 532 (1996).
\item V. S. Beskin and Yu. N. Pidoprygora, J. Exp. Theor. Phys.  {\bf 107}, 1025 (1995).
\item G. S. Bisnovatyi-Kogan, Physical Problems of the Stellar Evolution Theory (Nauka, Moscow, 1989) [in Russian].
\item G. S. Bisnovatyi-Kogan, Ya. M. Kazhdan, A. A. Klypin, et al., Sov. Astron. 23, 201 (1979).
\item A. Brandenburg and D. D. Sokoloff, Geophys. Astrophys. Fluid Dyn. {\bf 96}, 319 (2002).
{\bf 96}, 319 (2002).
\item K. G. Guderley, Theory of Transonic Flow (Pergamon, Oxford, 1962; Inostr. Liter., Moscow, 1960).
\item G. P. Horedt, Astron. Astrophys. {\bf 106}, 29 (1982).
\item R. Hunt, Mon. Not. R. Astron. Soc. {\bf 198}, 83 (1979).
\item J. Krolik and J. F. Hawley, Astrophys. J.  {\bf 573}, 754 (1991).
\item V. M. Lipunov, Astrophysics of Neutron Stars
(Nauka, Moscow, 1987; Springer, Heidelberg, 1992).
\item R. Mises, Mathematical Theory of Compressible Fluid Flow (Academic Press, New York, 1958; Inostr.
Liter., Moscow, 1961).
\item V. I. Pariev, Mon. Not. R. Astron. Soc. {\bf 283} 1264 (1996).
\item L. I. Petrich, S. Shapiro, and S. Teukolsky, Phys. Rev.
Lett.  {\bf 60}, 1781 (1988).
\item L. I. Petrich, S. Shapiro, R. F. Stark, and S. Teukolsky, Astropys. J.  {\bf 336} 313
(1989).
\item M. Ruffert and D. Arnett, Astron. Astrophys. {\bf 346}, 861 (1994).
\item N.I. Shakura, K.A. Postnov, A.Yu. Kochetkova and
L. Yalmasdotter, Phys. Usp. {\bf 183}, 337 (2013).
\item S. Shapiro and S. Teukolsky, Black Holes, White Dwarfs, and Neutron Stars: The Physics of Compact Objects (Wiley, New York, 1983; Mir, Moscow,
1985).
\item Yu. M. Toropin, O. D. Toropina, V. V. Saveliev, et al.,
Astorphys. J.  {\bf 593}, 472 (1999).
\item O. D. Toropina, M. M. Romanova, and
R. V. D. Lovelace, Mon. Not. R. Astron. Soc. {\bf 420}, 810 (2012).

\end{enumerate}

\end{document}